\ifx\onecol\undefined
\documentclass[journal,twocolumn]{IEEEtran}
\else 
\documentclass[12pt,draftclsnofoot,journal,onecolumn]{IEEEtran}
\fi
\usepackage{amsfonts}
\usepackage{amsmath,amssymb}
\usepackage{acronym}  
\usepackage{algorithm}
\usepackage{algorithmic}
\usepackage{bm}
\usepackage{bbm}
\usepackage{multirow}
\usepackage{booktabs}
\usepackage{color, soul}
\usepackage{cite}
\usepackage{graphicx}
\usepackage{indentfirst}
\usepackage{lipsum}
\usepackage{setspace}
\usepackage{tikz}
\usetikzlibrary{arrows}
\usepackage{subfigure}
\usepackage[amsmath,thmmarks]{ntheorem}
\usepackage{theorem}
\usepackage{amsmath}
\usepackage{url}
\usepackage{hyperref}

\begin{document}

\title{MUSE-FM: Multi-task Environment-aware Foundation Model for Wireless Communications}

\author{Tianyue Zheng, Jiajia Guo, {\textit{Member, IEEE}}, Linglong Dai, {\textit{Fellow, IEEE}} \\
Shi Jin, {\textit{Fellow, IEEE}}, and Jun Zhang, {\textit{Fellow, IEEE}} \vspace{-5mm}
\thanks{T. Zheng and L. Dai are with the Department of Electronic Engineering, Tsinghua University, and the State Key laboratory of Space Network and Communications, Tsinghua University, Beijing 100084, China (e-mails: zhengty22@mails.tsinghua.edu.cn, daill@tsinghua.edu.cn).

J. Guo and J. Zhang are with the Department of Electronic and Computer Engineering, The Hong Kong University of Science and Technology, Hong Kong (e-mails: eejiajiaguo@ust.hk, eejzhang@ust.hk).

S. Jin is with the National Mobile Communications Research Laboratory, Southeast University, Nanjing 210096, China (e-mails: jinshi@seu.edu.cn).
}} 

\maketitle

\begin{abstract}
Recent advancements in foundation models (FMs) have attracted increasing attention in the wireless communication domain. Leveraging the powerful multi-task learning capability, FMs hold the promise of unifying multiple tasks of wireless communication with a single framework. 
Nevertheless, existing wireless FMs face 
limitations in the uniformity to address multiple tasks with diverse inputs/outputs across different communication scenarios.
In this paper, we propose a MUlti-taSk Environment-aware FM (MUSE-FM) with a unified architecture to handle multiple tasks in wireless communications, while effectively incorporating scenario information. 
Specifically, to achieve task uniformity, we propose a unified prompt-guided data encoder-decoder pair to handle data with heterogeneous formats and distributions across different tasks. 
Besides, we integrate the environmental context as a multi-modal input, which serves as prior knowledge of environment and channel distributions and facilitates cross-scenario feature extraction.
Simulation results illustrate that the proposed MUSE-FM outperforms existing methods for various tasks, and its prompt-guided encoder-decoder pair facilitates few-shot adaptation to new task configurations. Moreover, the incorporation of environment information improves the ability to adapt to different scenarios.
\end{abstract}
\vspace{-2mm}

\vspace{-4mm}
\section{Introduction} \label{sec-intro}
\vspace{-1mm}
The integration of artificial intelligence (AI) with wireless communication has become increasingly vital for the design and optimization of sixth-generation (6G) wireless communication networks~\cite{6G2021,AI6G}.
Specifically, the integration of AI demonstrates performance improvement or complexity reduction for various communication tasks, including channel estimation and prediction, channel state information (CSI) feedback, channel decoding, and resource allocation under dynamic conditions~\cite{AI6G2}.
The recent paradigm shift towards Foundation Models (FMs), characterized by a large amount of parameters and pretraining on vast datasets, has dramatically reshaped the AI landscape~\cite{gpt}. 
These models exhibit remarkable reasoning capabilities and generalizability across multiple domains and tasks. 
Given their profound impact across diverse domains, exploring the applicability and potential of FMs in 6G wireless communications emerges as a crucial research imperative to unlock next-generation wireless intelligence.

\vspace{-3mm}
\subsection{Prior Works}
\vspace{-1mm}
There is an emerging trend of leveraging FMs to support massive device density, high-data-rate communications, and ubiquitous intelligence in 6G networks~\cite{LAM4PHY,LLM5,LLM6}.
To be specific,
existing works mainly have attempted to harness three key advantages of FMs, namely powerful feature extraction, few-shot learning, and generalization capability, to address challenges in wireless communication systems.
Firstly, owing to the substantial parameters, FMs acquire superior feature extraction capabilities compared to traditional deep learning (DL) architectures. This inherent advantage has attracted a growing research focus on leveraging these models to address complex problems and scenarios for various tasks in wireless communications, such as channel prediction and scatter generation.
Specifically, Liu et al.~\cite{LLM4CP} introduce FM to physical layer communications, and propose LLM4CP to deal with the challenge of channel prediction in high-dynamic scenarios. LLM4CP unleashes the power of FMs to achieve accurate channel prediction.
It is proposed in~\cite{ISAC} that an FM-enabled decomposition-based multi-objective evolutionary algorithm can jointly optimize the communication and sensing objectives in integrated sensing and communication (ISAC) systems.
\cite{CNN-GPT2} focuses on beam prediction task for near-field communications, which demonstrates the advantages of FM in handling complex temporal dynamics due to the joint dependence on user angle and distance for near field channels.
Besides, FMs are also employed in fluid antenna systems~\cite{FAS} to tackle the challenges of user mobility.

Secondly, pretrained on extensive datasets, FMs possess inherent few-shot/zero-shot learning capabilities, compared to conventional DL-based models.
The artificial general intelligence acquired by the FMs through pretraining enables direct task inference without the need to design and train dedicated networks from scratch. 
For example, in~\cite{power}, the authors adopt a pretrained FM to perform power allocation using a few-shot learning approach. In this method, the channel gains and the corresponding transmit power strategies are contacted as prompts and fed to the FM. Then, the FM can provide the power allocation strategy without any fine-tuning.
Thirdly, FMs exhibit better generalization capability to different wireless environments, compared to traditional DL models. 
For instance, leveraging this generalization capability, Guo et al.~\cite{LLMCSIFeed} propose an FM-based CSI feedback method that is adaptable to various scenarios. The authors incorporate the channel distribution in different environments as input and enable the model to handle multiple scenarios simultaneously.

It has been observed that FMs have successfully improved the performance and robustness in various wireless communication tasks~\cite{LLM4CP,ISAC,CNN-GPT2,FAS,power,LLMCSIFeed}.
Nevertheless, existing works typically design a dedicated FM for each individual task. Consequently, developing separate FMs for all the aforementioned tasks would incur substantial storage and deployment overhead.
Recently, a few initial attempts have been made to propose a multi-task FM to unify multiple tasks with a single model. 
In~\cite{LWM}, the authors propose an FM to extract features of wireless channels, which are used as common inputs of downstream models in downstream tasks.
There are also studies ~\cite{WirelessGPT,LLM4WM} focusing on channel-associated tasks such as channel estimation, beam selection and path loss estimation. They employ an FM to jointly learn a unified representation for multiple tasks, with task-specific adapters or fine-tuning for different downstream tasks. The authors in~\cite{MTLLM} attempt to integrate tasks with different data formats and distributions to fully demonstrate the multitasking capability of FMs. For this purpose, a multi-task FM is proposed to unify three tasks, namely channel prediction, signal detection and multi-user precoding, with the aid of task-specific adapters and elaborately designed prompts.

\vspace{-3mm}
\subsection{Motivations}
\vspace{-1mm}
Despite preliminary advances, existing wireless FMs face limitations in uniformity in addressing multiple tasks with diverse inputs/outputs in different communication scenarios.

\textbf{Limitation in task uniformity:}
Existing works suffer from the narrow task scope, as they predominantly focus on channel-related tasks.
These works utilize pilot signals or CSI as input, employing FMs to extract features for multiple downstream tasks.
However, they fail to unify other critical components across the entire physical layer functionalities. 
Besides, existing works employ task-specific data encoders and decoders to handle diverse input/output formats and feature spaces across different tasks and parameters.
However, given the diverse tasks and parameter variations, maintaining separate data encoder-decoder pairs for each task and parameter incurs high computational and deployment overhead. 
It also degrades the scalability of the model, contradicting the objectives of unified multi-task FM paradigms.

\textbf{Limitation in scenario uniformity:}
Scenario uniformity refers to the environmental adaptability of the model to generalize across multiple scenarios, which is critical for real-world deployment.
However, existing multi-task FMs are typically optimized for a specific scene. When encountered a new scenario with a different spatial structure, the shift in distribution of channel characteristics leads to severe performance distortion. 
In this case, data samples in the new environment are required to be collected and online fine-tuning is then performed, which imposes substantial computational overheads especially for FMs with large-scale parameters.
Therefore, existing works significantly compromise the environmental adaptability of FMs, and increase deployment costs.

Furthermore, under \textbf{data scarcity}, task-specific wireless FMs trained on limited single-task data are prone to overfitting, resulting in performance drop~\cite{LLM4WM}. This also motivates the design of multi-task wireless FMs that leverage intertask relationships to extract knowledge from diverse datasets, enabling superior joint optimization.


\vspace{-3mm}
\subsection{Our Contributions}
\vspace{-1mm}
To address the aforementioned limitations, in this work, we propose a MUlti-taSk Environment-aware FM (MUSE-FM) for wireless communications with a unified architecture to handle multiple tasks and incorporate scenario information. 
Specifically, the contributions are summarized as follows.
\begin{itemize}
    \item A multi-task wireless FM is proposed to enhance the uniformity to address different tasks with heterogeneous data formats and distributions. It is achieved by a unified prompt-guided data encoder-decoder pair.
    To be specific, the parameters of the prompt-guided unified data encoder-decoder pair are dynamically generated by task-specific instructions through hypernetworks, allowing instruction to ``guide” the feature extraction of data encoders (or output generation of data decoders).
    \item In addition, to improve the environmental uniformity, we integrate the environmental context as a multi-modal input. Specifically, prior knowledge of environment and channel distributions is obtained from scene graphs and facilitates cross-scenario feature extraction.
    Thus, we establish a multi-task environment-aware FM that dynamically adapts to varying propagation environments.
    \item A multi-task multi-scenario multi-modal dataset for physical layer communications is constructed to verify the effectiveness of the proposed method.
    The dataset contains 2600 indoor scenarios with different layouts, in each of which 50 data samples are generated. 
    The samples involve environmental information and data for various tasks for physical layer communications. 
    \item Extensive experiments have been conducted to evaluate the performance of the proposed method. It outperforms baselines for various tasks.
    Besides, incorporating environmental information facilitates cross-scenario learning ability. 
    The proposed prompt-guided unified encoder-decoder pair improves the scalability of the model while achieving comparable performance with task-specific encoder-decoder pairs.
    Furthermore, with limited data samples, MUSE-FM harnesses cross-task dependencies for knowledge transfer and enhances joint optimization compared with task-specific FMs.
\end{itemize}

The rest of the paper is organized as follows. Section II introduces the system model. We illustrate the typical structure of MIMO-OFDM transceivers and identify the essential yet challenging tasks in the transceivers. Then, the selected tasks are formulated, respectively. In Section III, we elaborate on the overall framework and specific design of the proposed MUSE-FM. The loss function and the training schedule are then specified.
Simulation results are provided in Section IV. Finally, Section V concludes this paper.

{\it Notation:} ${\bf a}^T$, ${\bf a}^H$ denote the transpose,  conjugate transpose of $\bf a$, respectively; $|{\bf a}|$ denotes the absolute value of $\bf a$ while $\left\| {\bf a} \right\|_2$ denotes the $l_2$ norm of $\bf a$; $\mathbb{R}$,$\mathbb{C}$ denote the set of real numbers and complex numbers, respectively; $\mathcal{C} \mathcal{N} ({\bf \mu},{\bf \Sigma})$ denotes the probability density function of complex multivariate Gaussian distribution with mean ${\bf \mu}$ and variance ${\bf \Sigma}$.

\vspace{-3mm}
\section{Systems Model}\label{sec-sys}
A multi-user (MU) multiple-input-single-output (MISO)- orthogonal-frequency-division-multiplexing (OFDM) system operating at mmWave frequency with $M$ subcarriers is considered, where a base station (BS) simultaneously serves $K$ users (UEs). The BS adopts $N_{t}=N_h \times N_v$ antennas arranged in a uniform planar array (UPA) configuration, where $N_h$ and $N_v$ denote the quantity of antennas along the horizontal and vertical dimensions, respectively. UEs are equipped with an omnidirectional antenna and can accommodate multiple antennas through parallel processing. 
In this work, we aim to deploy an FM at the BS to simultaneously handle most of the basic tasks for the transceiver. In the following, we first present a brief description of the typical structure of MIMO-OFDM transceivers and identify the basic and challenging tasks that require the employment of FM. Secondly, we present the problem formulations of the selected tasks sequentially.

\begin{figure*}
	\centering 
	\includegraphics[width= 0.9 \linewidth]{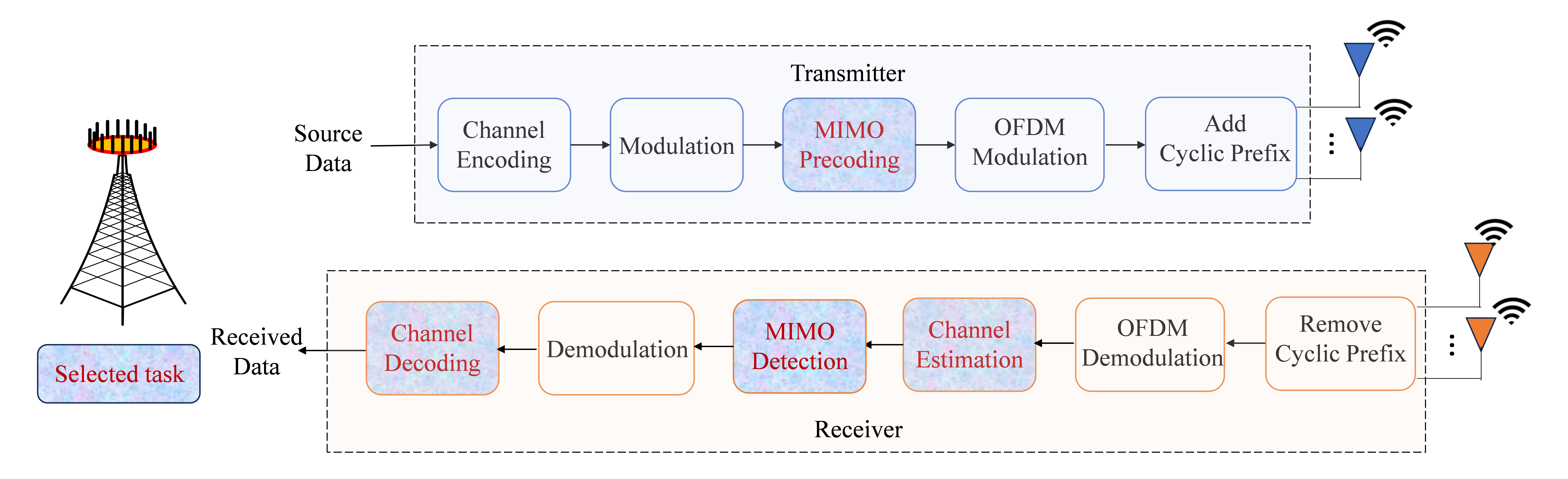}
        \caption{Typical structure of MIMO-OFDM transmitter and receiver.}
	\label{pic_transceiver}
    \vspace{-6mm}
\end{figure*}

\vspace{-3mm}
\subsection{The Structure of MIMO-OFDM Transceivers}
\vspace{-1mm}
The typical structure of a MIMO-OFDM transceiver, deployed at the BS, is depicted in Fig.~\ref{pic_transceiver}.
The MIMO-OFDM transmitter begins with channel encoding (e.g., LDPC or polar codes), which adds redundancy to the input bitstream $\mathbf{b}$ for robustness against channel errors. The encoded bits $\mathbf{s}$ are then mapped to complex symbols by modulation (e.g., QAM or PSK). For multi-user systems, multi-user precoding is applied to minimize inter-user interference and simultaneously transmit $\mathbf{x}_k, k=1,2,\cdots, K$ for $K$ users, based on the multi-user CSI. The modulated symbols are subsequently allocated to different OFDM subcarriers, added by a cyclic prefix (CP) to mitigate inter-symbol interference (ISI) caused by multi-path fading. 
The data transmitted to the $k$-th user of the $m$-th subcarrier, $m=1,2,\cdots,M$, is referred to as ${x}_{m,k}$. 

Due to the sparse scattering environment of the mmWave frequency~\cite{mmwave_channel}, we adopt the Saleh-Valenzuela channel Model for the multi-path channel $\mathbf{h}_{m,k} \in \mathbb{C}^{N_t \times 1}$ between the BS and the user $k$ at the $m$-th subcarrier as
\vspace{-1mm}
\begin{equation}
    \mathbf{h}_{m,k} = \sum_{l=1}^{L_k} \beta_{m,k,l} e^{-j2\pi f_m\tau_{k,l}} \mathbf{a}_m(\theta_{k,l}, \phi_{k,l}),
\vspace{-1mm}
\end{equation}
where $L_k$ is the number of paths, and $f_m$ is the frequency at the $m$-th subcarrier. $\beta_{m,k,l}$, $\tau_{k,l}$, $\theta_{k,l}$ and $\phi_{k,l}$ represent the complex path gain, delay, elevation angle, and azimuth angle of the $l$-th path of user $k$.
$\mathbf{a}_m(\theta, \phi)$ represents the steering vector of the corresponding path, which is derived for UPA as
\begin{align}
\vspace{-1mm}
    \mathbf{a}_m(\theta, \phi) 
    &= \frac{1}{\sqrt{N_t}} [ 1, \cdots, e^{j k_m d(n_v \sin\phi \sin\theta + n_h \cos\theta)}, \cdots, \nonumber \\ 
    \quad \quad \quad  \quad & e^{j k_m d((N_v - 1)\sin\phi \sin\theta + (N_h - 1)\cos\theta)} ]^T, 
\vspace{-1mm}
\end{align}
where $k_m = \frac{2\pi}{\lambda_m}$, $\lambda_m$, and $d$ denote the wavenumber, wavelength of the $m$-th subcarrier, and spacing of adjacent antenna elements, respectively.

Next, we illustrate the receiver of the BS. 
The BS antenna array simultaneously receives the transmitted symbols from $K$ users. 
Then CP is removed to eliminate ISI and OFDM demodulation is performed. During the pilot transmission, the BS estimates the channel $\hat{\mathbf{H}}$. Based on the estimated channel, MIMO detection is performed to recover the transmitted symbol $\hat{\mathbf{x}}$ during data transmission. The detected symbols undergo demodulation to recover the encoded bits $\hat{\mathbf{s}}$ according to the modulation scheme. Finally, channel decoding corrects transmission errors by exploiting redundancy from the encoder, restoring the original bitstream $\hat{\mathbf{b}}$. 

Building upon the aforementioned description of the MIMO-OFDM transceiver, we select critical yet challenging tasks including \textit{channel estimation}, \textit{MIMO detection}, \textit{channel decoding}, and \textit{multi-user precoding}, while ingoring the relatively trivial components (e.g., channel encoding, which requires only a matrix multiplication operation). The task selection aims to establish a unified FM capable of addressing the most demanding aspects of a transceiver. Furthermore, we incorporate \textit{user localization} as an essential task, given its fundamental role in 6G ISAC systems, a pivotal direction for next-generation wireless networks.
The designed multi-task FM aims to handle tasks across various physical layers as comprehensively as possible.
In the next subsection, we will formulate the selected tasks individually.

\vspace{-3mm}
\subsection{Problem Descriptions of Selected Tasks}
\vspace{-1mm}
\subsubsection{Channel estimation}
In MIMO-OFDM systems, the accuracy of channel estimation significantly affects the subsequent signal detection and demodulation process. To obtain CSI, pilots are transmitted and the corresponding received signal $\mathbf{Y}^p_k$ of the $k$-th user is written as
\begin{equation}
    \mathbf{Y}^p_k = {\mathbf{W}}_s {\mathbf{H}}_k + {\mathbf{N}}_k, 
\end{equation}
where ${\mathbf{H}}_k=[{\mathbf{h}}_{1,k},{\mathbf{h}}_{2,k},\cdots,{\mathbf{h}}_{M,k}] \in  \mathbb{C}^{N_{t} \times M}$ is the channel of the $M$ subcarriers for the user $k$,  ${\mathbf{W}}_s \in  \mathbb{C}^{L_p \times N_{t}}$ is the selection matrix at the BS with pilot length $L_p$, ${\mathbf{N}}_k$ is the additive Gaussian white noise (AWGN). A channel estimator is designed to construct the CSI ${\mathbf{H}}_k,\forall k$ from the received signal $\mathbf{Y}^p_k \in  \mathbb{C}^{L_p \times M}$, that is.
\begin{align}
\textbf{P1: } \min_{\Omega_{ce}} \quad & ||\hat{{\mathbf{H}}}_k - {\mathbf{H}}_k||_2,  \forall k \\
\text{s.t.} \quad & \hat{{\mathbf{H}}}_k = f_{\Omega_{ce}}(\mathbf{Y}^p_k),
\end{align}
where the estimated channel is obtained from the neural network with a mapping function $f_{\Omega_{ce}}$. The neural network takes $\mathbf{Y}^p_k$ as input, with learnable parameters ${\Omega_{ce}}$.

\subsubsection{MIMO detection}
During uplink data transmission, the BS antenna array simultaneously receives the symbols transmitted by the $K$ users. 
Specifically, denote the transmitted symbol vector by all users at the $m$-th subcarrier as $\mathbf{x}_{m} = [{x}_{m,1}, {x}_{m,2}, \cdots, {x}_{m,K}] \in \mathbb{C}^{K \times 1}$.
The received signal $\mathbf{y}_m \in \mathbb{C}^{N_{t} \times 1}$ is given by
\begin{equation}
    {\mathbf{y}}_m = {\mathbf{H}}_m {\mathbf{x}}_m + {\mathbf{n}}_m,
\end{equation}
where ${\mathbf{H}}_m = [{\mathbf{h}}_{m,1},{\mathbf{h}}_{m,2}.\cdots,{\mathbf{h}}_{m,K}]$ is the uplink channel of the $K$ users and ${\mathbf{n}}_m \sim \mathcal{CN}(0,{\sigma}^2 \mathbf{I}_{N_{\rm T}})$ is the AWGN at the $m$-th subcarrier.

The BS will recover the signals ${\mathbf{x}}_m$ from the received signals ${\mathbf{y}}_m$ and $\hat{\mathbf{H}}_m$. 
For the MIMO detection task, we can independently recover the signal of different subcarriers.
We adopt the minimum mean squared error (MMSE) estimator to formulate the associated MIMO detection problem as
\begin{align}
\textbf{P2: } \min_{\Omega_{det}} \quad & ||\hat{{\mathbf{x}}}_m - {\mathbf{x}}_m||_2,  \forall m \\
\text{s.t.} \quad & \hat{{\mathbf{x}}}_m = f_{\Omega_{det}}(\hat{\mathbf{H}}_m,{\mathbf{y}}_m),
\end{align}
where $f_{\Omega_{det}}$ is the mapping function with parameters $\Omega_{det}$.

\subsubsection{Multi-user precoding} 
Multi-user precoding aims to maximize the sum rate of $K$ users through the optimization of the transmit precoders, based on the multi-user channel. The total power of all precoding vectors is limited due to the BS power budget. For simplicity, we parallelize the precoding for different subcarriers. That is, we design the precoder of each subcarrier separately. Therefore, the problem is mathematically formulated as
\begin{align}
\textbf{P3: } \max_{\Omega_{precoding}} \quad & \sum_{k=1}^{K} \log_2(1 + \gamma_{m,k}), \forall m  \label{eq-rate}\\
\text{s.t.} \quad & \sum_{k=1}^{K} \|\mathbf{w}_{m,k}\|^2 \leq P_{\max}, \\ \quad \quad & {{\mathbf{W}}}_m = f_{\Omega_{precoding}}(\hat{\mathbf{H}}_m), 
\end{align}
where $\mathbf{W}_m = [\mathbf{w}_{m,1},\mathbf{w}_{m,2},\cdots,\mathbf{w}_{m,K}]$ is the designed precoders, $\hat{\mathbf{H}}_m=[\hat{\mathbf{h}}_{m,1},\hat{\mathbf{h}}_{m,2},\cdots,\hat{\mathbf{h}}_{m,K}]$ is the estimated channel of $K$ users, and $P_{\rm max}$ is the power budget.
The precoder $\mathbf{W}_m$ is learned from the neural network with the mapping function $f_{\Omega_{ precoding}}$, where $\Omega_{ precoding}$ denotes the variable parameters.
Besides, $\gamma_{m,k}$ represents the received signal-to-interference-plus-noise ratio (SINR) at user $k$:
\begin{equation}
    \gamma_{m,k} = \frac{\left| \mathbf{h}_{m,k}^{H} \mathbf{w}_{m,k} \right|^2}{\sum_{k'=1, k' \neq k}^{K} \left| \mathbf{h}_{m,k}^{H} \mathbf{w}_{m,k'} \right|^2 + \sigma^2},
\end{equation}
where $\sigma^2$ represents the variance of AWGN.

\subsubsection{Channel decoding}
Channel decoding aims to recover the information bitstream ${\mathbf{b}} \in \{0,1\}^m$ from the estimated encoded bitstream $\hat{\mathbf{s}}$, through a neural network with a parameterized mapping function $f_{\Omega_{decoding}}$.
The recovered $\hat{\mathbf{s}}$ can be modeled as $\hat{\mathbf{s}} = \mathbf{s} + n_s$, with $\mathbf{s} \in \{\pm 1\}^n$ and $n_s$ being noise.
To achieve efficient channel decoding and facilitate fair comparison, we follow the pre-processing, post-processing, and problem descriptions of ECCT~\cite{ECCT}.
To be specific, the pre-processing process replaces $\hat{\mathbf{s}}$ with a vector of dimensionality $2n-m$ defined as
\begin{equation}
    \tilde{{\mathbf{s}}} = [|\hat{\mathbf{s}}|,h(\hat{\mathbf{s}})],
\end{equation}
where $|\hat{\mathbf{s}}|$ signifies the absolute value of $\hat{\mathbf{s}}$. $h(\hat{\mathbf{s}})= \mathbf{P} \cdot {\rm bin(sign(\hat{\mathbf{s}}))} \in \{0,1\}^{n-m}$ is the syndrome from hard-decoded $\hat{\mathbf{s}}$, where $\mathbf{P} \in \mathbb{R}^{(n-m) \times n}$ is the parity check matrix.
Then $\tilde{{\mathbf{s}}}$ is fed to the neural network and the predicted noise $\hat{\mathbf{z}}$ is obtained (we suppose $\hat{\mathbf{s}} = \mathbf{s} \cdot \mathbf{z}$).
In the post-processing step, the predicted noise $\hat{\mathbf{z}}$ is multiplied by $\hat{\mathbf{s}}$ to recover ${\mathbf{b}}$. That is, the predicted $\hat{\mathbf{b}}$ takes the form:
\begin{equation}
    \hat{\mathbf{b}} = {\rm bin(sign(\hat{\mathbf{s}} \cdot \hat{\mathbf{z}}))}.
\end{equation}

We employ binary cross-entropy as the loss function with the objective of learning to predict multiplicative noise $\mathbf{z}$. The corresponding target binary multiplicative noise is denoted as $\tilde{\mathbf{z}} = {\rm bin(sign(\mathbf{z}))} = {\rm bin(sign(\mathbf{\hat{\mathbf{s}}} \cdot {\mathbf{s}}))}$.
Therefore, the problem can be modeled as
\begin{align}
\textbf{P4: } \min_{\Omega_{decodig}} \quad & - \sum_{i=1}^n \tilde{z}_i \log (\hat{{z}_i}) + (1 - \tilde{z}_i) \log (1 - \hat{{z}_i}), \label{eq-deloss} \\
\text{s.t.} \quad & \hat{{\mathbf{z}}} = f_{\Omega_{decoding}}(\tilde{{\mathbf{s}}}),
\end{align}
where $\tilde{z}_i$ and $\hat{{z}_i}$ are the $i$-th element of $\tilde{\mathbf{z}}$ and $\hat{\mathbf{z}}$, respectively.

\subsubsection{User localization}
In 6G networks, ISAC emerges as a transformative paradigm, merging sensing and communication, where precise user localization is essential to enable dynamic resource allocation and intelligent network control. Therefore, in this paper, we not only attempt to estimate the channel from $\mathbf{Y}^p_k$, but also learn the user positions to facilitate subsequent ISAC applications.
Similar to channel estimation, the user localization problem can be formulated as
\begin{align}
\textbf{P5: } \min_{\Omega_{loc}} \quad & ||\hat{{\mathbf{pos}}}_k - {\mathbf{pos}}_k||_2,  \forall k \\
\text{s.t.} \quad & \hat{{\mathbf{pos}}}_k = f_{\Omega_{loc}}(\mathbf{Y}^p_k),
\end{align}
where ${\mathbf{pos}}_k$ and $\hat{{\mathbf{pos}}}_k$ represent the actual and estimated position, respectively. Besides, $\Omega_{loc}$ is the trainable parameters of the mapping function $f_{\Omega_{loc}}$.

\section{Multi-task Environment-aware Foundation Model (MUSE-FM)}\label{sec-pro}
In this section, we first introduce the overall framework of the proposed multi-task environment-aware FM (MUSE-FM). Then, the specific designs of the components in the framework are elaborated, respectively. Finally, the loss function and training schedule are illustrated.

\begin{figure*}
	\centering 
	\includegraphics[width= 0.9 \linewidth]{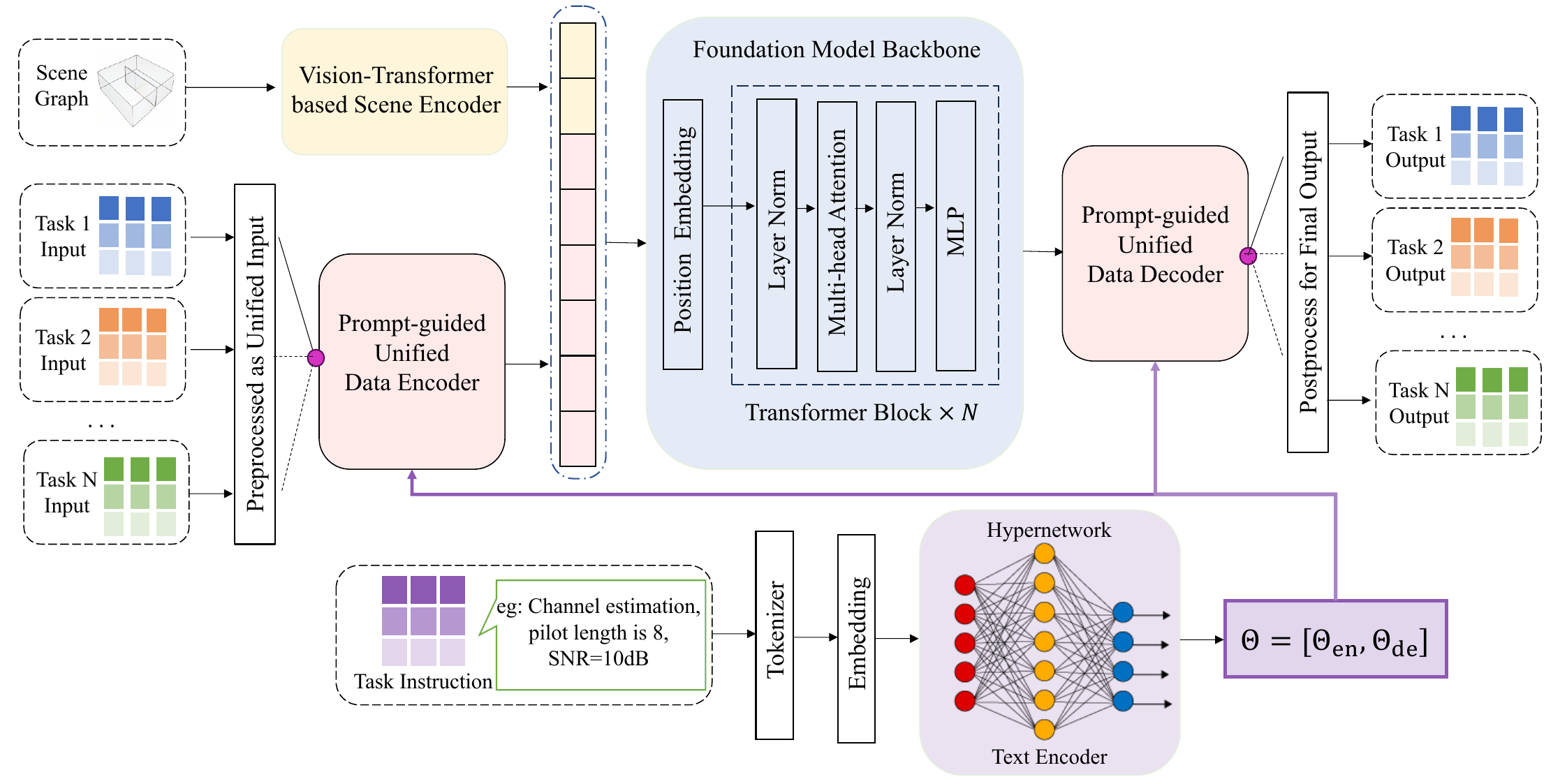}
    \vspace{-4mm}
    \caption{The model framework of our method.}
	\label{pic_frame}
    \vspace{-3mm}
\end{figure*}

\begin{figure}
\vspace{-3mm}
	\centering 
	\includegraphics[width= 0.95 \linewidth]{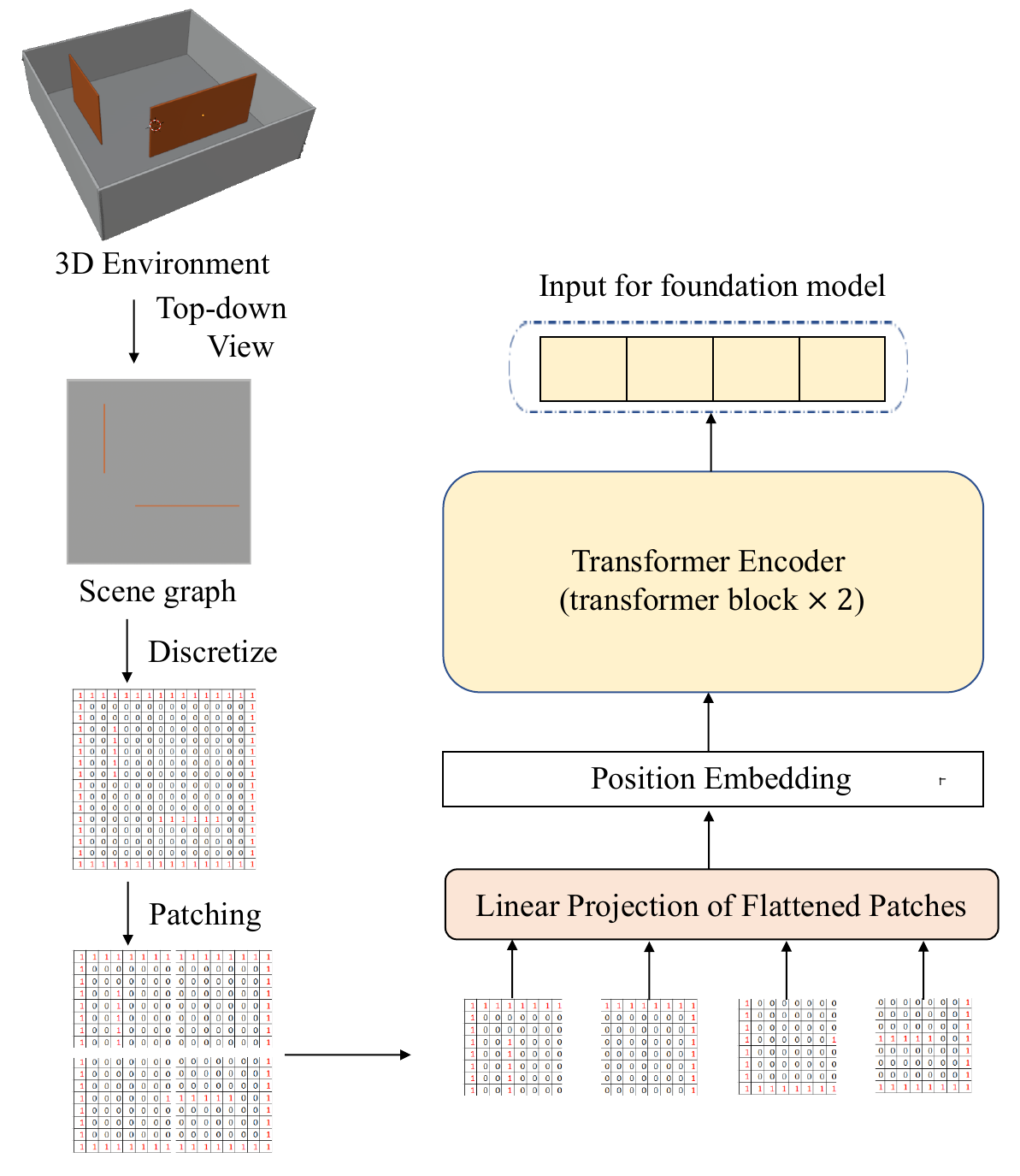}
    \vspace{-4mm}
        \caption{Vision-transformer based scene encoder.}
	\label{pic_scene}
    \vspace{-5mm}
\end{figure}

\vspace{-3mm}
\subsection{Overall Framework}\label{sec3-a} 
\vspace{-1mm}
The framework of the proposed model is presented in Fig.~\ref{pic_frame}, which includes the scene encoder, prompted-guided unified encoder-decoder pair, and FM backbone.
To address scenario limitation, leveraging the strong correlation between physical environments and channel characteristics~\cite{ChannelGPT}, we attempt to incorporate environmental information to allow cross-scenario generalization in multi-task FM.
Specifically, the spatial structure of the physical environment, including the area building distribution as well as the BS position, determines the propagation path information of the wireless channel, and thus directly impacts channel characteristics. Therefore, involving the environmental knowledge as a multi-modal input provides a promising way to inform FMs of the channel distribution and improves the generalization ability in different scenarios.
We implement it through a Vision Transformer-based encoder that extracts environmental and channel prior knowledge from scene graphs. The task-agnostic prior knowledge is subsequently fused with wireless data to enhance the model's environmental adaptability.
The specific design of the scene encoder is described in Section~\ref{sec3-b}.

To handle data with heterogeneous formats and distributions across multiple tasks, we propose a prompt-guided unified data encoder whose parameters are dynamically generated by task-specific instructions through hypernetworks. 
In particular, the task instruction is fed to a hypernetwork to dynamically generate parameters, which is utilized in the data encoder.
This architecture allows the instruction to ``guide'' the encoder to adaptively extract task-specific features according to each task's objectives and data characteristics.
To achieve this, two key components are included: \textbf{a parameter generator}, i.e., which is used to generate task-specific parameters from task instruction via a text encoder; \textbf{a prompted-guided unified encoder-decoder pair}, to utilize the generated parameters to project the data of multiple tasks into the shared FM space (or generate the final outputs in multiple tasks). The two components are elaborated in Section~\ref{sec3-c} and~\ref{sec3-d}.

The token embeddings from scene graphs and wireless data (e.g., CSI and received symbols) are concatenated to form the input embeddings for the FM backbone. 
The FM backbone consists of position embedding and stacked classical transformer blocks, which is illustrated in Section~\ref{sec3-e}.
Equipped with large-scale parameters and a layered structure, the FM backbone exhibits expressive learning capacity for multiple tasks.
Besides, atthention mechanism enables effective fusion of multi-modal inputs. It assimilates related channel features from task-agnostic prior knowledge for different tasks and integrates it with wireless data, ensuring cross-scenario feature extraction.
Finally, similar to the data encoder, equipped with task-aware parameters, the prompt-guided unified decoder generates the output for different tasks. 

MUSE-FM is designed to handle a wide range of physical-layer tasks. When encountering novel tasks, its uniform architecture allows for extension through minimal task-specific adaptation by designing a dedicated task instruction and fine-tuning the model with limited data samples.

\vspace{-3mm}
\subsection{Scene Encoder}\label{sec3-b}
\vspace{-1mm}
The scene encoder aims to extract task-agnostic priors of channel and environment from scene graphs.
The structure of the scene encoder to effectively exploit environmental information is illustrated in Fig.~\ref{pic_scene}. 
Direct utilization of raw 3D environmental data is challenging due to its high dimensionality. We therefore implement \textbf{preprocessing}, transforming environmental information into compact representations suitable for encoder input. Given that top-down views holistically encapsulate the spatial structure of the environment, that is, scene layouts and obstacle distributions, we extract it to construct the scene graph~\cite{liu2025adapcsinet}. This scene graph is then discretized into a matrix form, where the value 1 represents the obstacles, and the value 0 represents free space.

Obtaining the discretized scene graph, denoted by $\mathcal{G}\in \mathbb{R}^{W\times W}$, a \textbf{vision transformer-based scene encoder} is performed to extract prior channel knowledge from the physical environment. 
Specifically, to handle the 2D scene graph, we reshape $\mathcal{G}$ into a sequence of flattened patches $\mathcal{G}_p \in \mathbb{R}^{L_s \times P^2}$, where $(P, P)$ is the resolution of each image patch, and $L_s = W^2/P^2$ is the resulting number of patches. Then a trainable linear projection is utilized to map the flattened patches to the feature dimension of transformer encoder $D$. We refer to the output of this projection as the patch embeddings $\mathrm{X}_s^{\rm patch} \in \mathbb{R}^{L_s\times D}$:
\begin{equation}
    \mathrm{X}_s^{\rm patch} = {\rm Linear}(\mathcal{G}_p).
\end{equation}
Position embeddings are added to patch embeddings to retain positional information, and the resulting sequence of embedding $\mathrm{X}_s^{\rm input}$ serves as input to the transformer encoder. The Transformer encoder~\cite{transformer} consists of alternating layers of multi-head self-attention and multi-layer perceptron (MLP) blocks. Layer normalization (LN) is applied before every block, and residual connections are added after every block. The transformer encoder is summarized as 
\begin{equation}
    \mathrm{X}_s^{\rm emb} = {\rm Transformer}(\mathrm{X}_s^{\rm input}),
\end{equation}
where $\mathrm{X}_s^{\rm emb}$ is then fed to the FM as part of the input.

\vspace{-4mm}
\subsection{Parameter Generator with Task Instruction}\label{sec3-c}
\vspace{-1mm}
\subsubsection{Design of task instruction}
Concise and informative task instruction is essential to generate task-related parameters and thus ``guide" the feature extraction of the data encoder and decoder. The designed task instruction includes two parts: \textbf{a task identifier}, and \textbf{key parameters}.
The \textbf{task identifier} provides a distinct identifier for each task. In this work, the task identifier includes ``Channel estimation", ``MIMO detection", ``Channel decoding", ``Multi-user precoding", and ``User localization".
\textbf{Key parameters} aim to involve important domain knowledge as priors to facilitate better understanding of the task. Taking channel estimation as an example, the pilot length and SNR of the received signal are incorporated, i.e., ``pilot length is x, SNR = x dB". 
\footnote{The designed task instructions for other tasks are ``MIMO detection, transmitting antenna number is x, data length is x, SNR is x dB" for MIMO detection; ``Channel decoding for polar codes with encoded bit length n = x and information bit length m = x" for channel decoding; ``Multi-user precoding, user number is x, SNR is x dB" for multi-user precoding; ``User localization, signal length for localization is x, SNR is x dB” for user localization.}

\subsubsection{Hypernetwork}
A hypernetwork is designed to generate parameters for another neural network. In this work, we employ a hypernetwork-based architecture~\cite{hypernetworks}, which learns a mapping function that dynamically produces task-related parameters for the data encoder and decoder. By incorporating task instructions, the approach allows the data encoder-decoder pair to automatically adapt to diverse communication tasks and parameters. The specific designs are illustrated as follows. 

Firstly, the task instruction is tokenized into vocabulary indices by the FM tokenizer (for example, we employ the pretrained tokenizer of GPT2). Then the vocabulary indices are mapped to high-dimensional token embeddings. The token embeddings, denoted as $\mathrm{X}_t^{\rm emb}$, serve as the input of the hypernetwork. The hypernetwork is an MLP module, consisting of alternating linear layers and ReLU activation functions. The generated task-aware parameters for the data encoder-decoder pair are written as:
\begin{equation}
    \Theta = [\Theta_{en},\Theta_{de}] = {\rm MLP}(\mathrm{X}_t^{\rm emb}),
\end{equation}
where $\Theta_{en}$ and $\Theta_{de}$ are respectively the parameters for data encoder and data decoder. 

\vspace{-2mm}
\subsection{Prompted-guided Unified Data Encoder and Decoder}\label{sec3-d}
After the parameter generator produces the task-specific parameters, we focus on how the parameters can be used in the unified encoder and decoder.

\subsubsection{Pre-processor and unified data encoder}
To enable a unified data encoder to process multi-task data with different formats and distributions, preprocessing is required for each task to preprocess the input data.
Since the transformer architecture inherently acquire the capability to process variable-length sequences, preprocessing primarily aims to standardize input feature dimensions via methods like zero-padding and unify distribution through batch normalization. Here, we process the input data to the same feature dimension $2N_t$ and the preprocessing process for the task $n$ can be expressed as:
\begin{equation}
    \mathrm{X}_n^{\rm pre} = {\rm Preprocessor}(\mathrm{X}_n^{\rm input}),
\end{equation}
where $n$ denotes the task identifier, and $\mathrm{X}_n^{\rm pre}$ and $\mathrm{X}_n^{\rm input}$ denote the preprocessed input and the original input, respectively.

For MIMO detection and multi-user precoding, the feature dimension of input real tensors is exactly $2N_t$ (the input of multi-user precoding is multi-user channel $\mathrm{X}_{precoding}^{\rm input} = \hat{\mathbf{H}}_m  \in \mathbb{R}^{K\times 2N_t}, \forall m$; the input of MIMO detection is the concatenation of the multi-user channel and the received data $\mathrm{X}_{det}^{\rm input} = [\hat{\mathbf{H}}_m,\mathbf{Y}_m]\in \mathbb{R}^{(K+L_d)\times 2N_t}$, where $L_d$ is the data length, $\mathbf{Y}_m = [\mathbf{y}_m^1,\mathbf{y}_m^2,\cdots,\mathbf{y}_m^{L_d}]$ is the  concatenated received data of $L_d$ time slots. Thus, batch normalization can be directly performed for the input data to facilitate convergence of network training. For channel estimation and user localization, the input data is received OFDM signal of pilot $\mathrm{X}_{ce}^{\rm input} = \mathrm{X}_{loc}^{\rm input} = \mathbf{Y}_k^p \in \mathbb{R}^{M \times 2L_p}, \forall k$, where $L_p$ denotes pilot length. Thus, zero-padding is required to align feature dimension $2N_t$ and then batch normalization is performed. Finally, for channel decoding the one-dimensional input signal $\mathrm{X}_{decoding}^{\rm input} = \tilde{{\mathbf{s}}} \in \mathbb{R}^{2n-m}$ is converted to $\mathrm{X}_{decoding}^{\rm pre} = diag(\mathrm{X}_{decoding}^{\rm input})$, where $diag(\mathrm{s})$ is a diagonal matrix with $\mathrm{s}$ on the diagonal.

\begin{figure}
	\centering 
	\includegraphics[width= \linewidth]{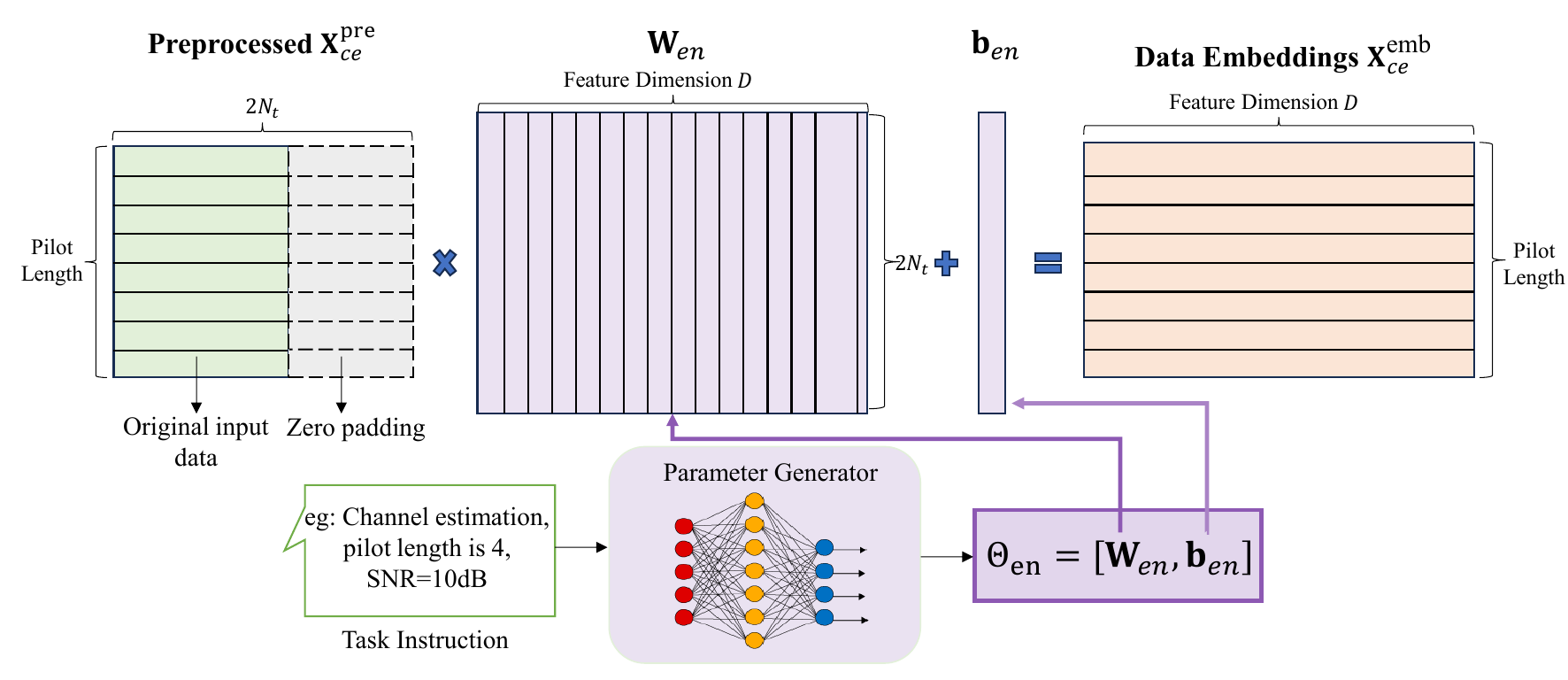}
        \caption{Illustration of the unified data encoder.}
	\label{pic_example}
\vspace{-3mm}
\end{figure}

After preprocessing, $\mathrm{X}_n^{\rm pre}$ is fed into the unified encoder. Specifically, the generated parameters $\Theta_{en}$ consist of the weight matrix $\mathrm{W}_{en} \in \mathbb{R}^{D\times 2N_t}$ and the bias vector $\mathrm{b}_{en} \in \mathbb{R}^{D\times 1}$. The parameters are applied to project $\mathrm{X}_n^{\rm pre}$ of different  tasks to a shared foundation model feature space:
\begin{equation}
    \mathrm{X}_n^{\rm emb} =  {\rm Linear}(\mathrm{X}_n^{\rm pre};\mathrm{W}_{en},\mathrm{b}_{en}) .
\end{equation}
In Fig.~\ref{pic_example}, we take channel estimation as an example to illustrate how the original data input $\mathrm{X}_n^{\rm pre}$ is preprocessed and handled by the unified data encoder, obtaining $\mathrm{X}_n^{\rm emb}$.

Finnaly, the data embeddings $\mathrm{X}_n^{\rm emb}$ are concatenated with the scene embeddings $\mathrm{X}_s^{\rm emb}$. 
Besides, we use the standard approach of adding an extra learnable “classification token” (cls token) to the sequence, to form the final input of the FM backbone. We refer to the input as $\mathrm{X}_{\rm FM}^{\rm emb}$.

\subsubsection{Post-processor and unified data decoder}
Obtaining the output of the FM backbone, denoted as $\mathrm{X}_{\rm FM}^{O}$, the unified data decoder and the post-processor are employed to generate the final output.
The parameters $\Theta_{de}$ capture the requirements of the corresponding task, enabling the data decoder to adaptively output the results for different tasks.
Similarly, $\Theta_{de}$ includes the weight matrix $\mathrm{W}_{de} \in \mathbb{R}^{2N_t \times D}$ and the bias vector $\mathrm{b}_{de} \in \mathbb{R}^{2N_t \times 1}$. We first discard the prefix portion that is associated with the scene, while we only retain the output presentations of the data, $\mathrm{X}_n^{\rm de}$, for the decoder. The output of the data decoder is written as
\begin{equation}
    \mathrm{X}_n^{\rm post} = {\rm Linear}(\mathrm{X}_n^{\rm de};  \mathrm{W}_{de}, \mathrm{b}_{de}).
\end{equation}

Finally, the post-processor extracts the corresponding output for the target task, which can be expressed as
\begin{equation}
    \mathrm{X}_n^{\rm output} = {\rm Postprocessor}(\mathrm{X}_n^{\rm post}).
\end{equation}
For channel estimation and multi-user precoding, the resulted $\mathbf{X}_{ce}^{\rm post}$ and $\mathbf{X}_{precoding}^{\rm post}$ have the same feature dimension with required final outputs. Thus, $\mathbf{X}_{ce}^{\rm output} = \mathbf{X}_{ce}^{\rm post}[1:,:] = \hat{\mathbf{H}}_k$ is the estimated channel, while $\mathbf{X}_{precoding}^{\rm output} = \mathbf{X}_{precoding}^{\rm post}[1:,:] = {\mathbf{W}}_m$ is the designed beamforming vector. For user localization, we extract the estimated user localization from the cls token, i.e. $\mathbf{X}_{loc}^{\rm output} = \mathbf{X}_{loc}^{\rm post}[0,:2] = \hat{\mathbf{pos}}$. For MIMO detection, the recovered data is written as $\mathbf{X}_{det}^{\rm output} = \mathbf{X}_{det}^{\rm post}[-L_d:,:2K] = \hat{\mathbf{x}}_m$. Lastly, for channel decoding, the recovered bit stream is presented as $\mathbf{X}_{decoding}^{\rm output} = \mathbf{X}_{decoding}^{\rm post}[:,0] = \hat{\mathbf{b}}$.

\subsection{Foundation Model Backbone}\label{sec3-e}
\vspace{-1mm}
The FM backbone fuses multi-modal inputs and effectively extracts features for multi-task learning. Leveraging the Transformer's attention mechanism, the FM dynamically assigns varying importance to tokens representing different environmental features. This enables it to distill task-beneficial representations from a unified prior knowledge of channel distribution $\mathrm{X}_s^{\rm emb}$ for multiple tasks.

Architecturally, it is composed of stacked Transformer modules, which is elaborated in detail.
Position embeddings are added to the combined tokens $\mathrm{X}_{\rm FM}^{\rm emb}$, providing sequential information.
Thus, the input to the transformer blocks is denoted as $\mathrm{X}_{\rm FM}^{I(1)}$.
As described in the scene encoder, each transformer block has two sub-layers. The first is a multi-head self-attention mechanism, and the second is a simple MLP network. 
The $l$-th transformer block takes the output of the $l-1$-th transformer block as input, that is, $\mathrm{X}_{\rm FM}^{I(l)} = \mathrm{X}_{\rm FM}^{O(l-1)}$. LN operates independently across the feature dimension for each token in the sequence and each sample in the batch:
\begin{equation}
    \mathrm{X}_{\rm FM}^{ln1(l)} = {\rm LayerNorm}(\mathrm{X}_{\rm FM}^{I(l)}), 
\end{equation}
where ${\rm LayerNorm}(\cdot)$ denotes the LN.
Then $\mathrm{X}_{\rm FM}^{ln1(l)}$ is processed by the multi-head self-attention module with residual connection as
\begin{equation}\label{eq-att}
    \mathbf{X}_{\rm FM}^{att(l)} = {\rm ATT}(\mathbf{X}_{\rm FM}^{ln1(l)}) + \mathbf{X}_{\rm FM}^{ln1(l)},
\end{equation}
where ${\rm ATT}(\cdot)$ denotes multi-head self-attention.
Then another LN is applied, obtaining $\mathrm{X}_{\rm FM}^{ln2(l)}$, which is processed by the MLP module, that is
\begin{equation}
    \mathbf{X}_{\rm FM}^{O(l)} = {\rm MLP}(\mathbf{X}_{\rm FM}^{ln2(l)}) + \mathbf{X}_{\rm FM}^{ln2(l)},
\end{equation}
where ${\rm MLP}(\cdot)$ is the MLP network and $\mathbf{X}_{\rm FM}^{O(l)}$ the output of the $l$-th transformer block.
After progressively extracting and integrating features with $L$ transformer blocks, the output of the FM is denoted as $\mathbf{X}_{\rm FM}^{O}.$ 

\vspace{-3mm}
\subsection{Loss Function and Training Schedule}\label{sec4-c}
The proposed MUSE-FM is trained on a mixed dataset with multiple tasks in different scenarios.
The multi-task loss function for the training stage is written as 
\begin{equation}
    {\rm Loss}_{train} = \sum_n \alpha_n {\rm loss}_{train,n},
\end{equation}
where ${\rm loss}_n$ is the loss function of task $n$. 
Besides, due to the disparities in the convergence rates and loss scales across tasks, we implement a weighted linear combination scheme with $\alpha_n$.
Specifically, $\alpha_n$ are determined with Dynamic Weight Average (DWA) algorithm~\cite{DWA}, to dynamically adjust each task's weight based on its loss every epoch.
Next, we briefly illustrate the loss function for each task.
For multi-user precoding, we directly optimize the negative sum rate in an unsupervised manner, which is presented in~\eqref{eq-rate}.
For regression problems including channel estimation, user localization, and MIMO detection, the MSE loss is selected as the loss function. For channel decoding, following~\cite{ECCT}, we utilize the cross entropy loss to optimize the estimated binary multiplicative noise. The selected tasks, the corresponding training metrics and evaluation metrics are summarized in Table~\ref{tab:list}.

\begin{table}[t]
\centering
\caption{Selected tasks and the corresponding training/evaluation metrics.}
\label{tab:list}
\begin{tabular}{c|cc}
\toprule
\textbf{Task} & \textbf{Training metric} &  \textbf{Evaluation metric} \\
\midrule
    CE & MSE & NMSE \\
    Precoding & Negative sum rate & Sum rate \\
    Det  & MSE & NMSE \\
    Decoding & cross entropy & BER\\
    Loc & MSE  & Error distance\\
\bottomrule
\end{tabular}
\vspace{-3mm}
\end{table}

During the training process, the model with the smallest validation loss is saved for the testing phase. Similar to the training loss, the validation loss is the weighted sum of the losses of individual tasks, i.e.
\begin{equation}
    {\rm Loss}_{val} = \sum_n \beta_n {\rm loss}_{val,n}.
\end{equation}
where $\beta_n$ and ${\rm loss}_{val,n}$ denote the weight and loss of the task $n$, respectively. Specifically, NMSE is utilized to evaluate the performance of channel estimation and MIMO detection; while the normalized average positioning error is chosen to provide intuitive evaluation for user localization. For multi-user precoding, the loss function is the negative sum rate; for channel decoding, the loss function is the bit error ratio (BER).

\vspace{-3mm}
\section{Simulation Results}\label{sec-sim}
In this section, extensive numerical simulations are presented to verify the effectiveness of the proposed method. First, we introduce the constructed multi-task multi-scenario multi-modal dataset. The detailed simulation setup and training configurations are then illustrated. Finally, we provide thorough performance evaluation and analysis of MUSE-FM.

\vspace{-3mm}
\subsection{Dataset Construction}
To validate the proposed method, it is necessary to generate a multi-task dataset with corresponding environmental information across multiple scenarios. 
The constructed dataset consists of data collected in 2600 scenarios; for each scenario, 50 data samples are simulated.
The dataset is partitioned into three subsets: 2000 scenarios for training, 300 scenarios for validation, and 300 scenarios for testing.

\subsubsection{Environmental information}
\begin{figure}
	\centering 
	\includegraphics[width= 0.85 \linewidth]{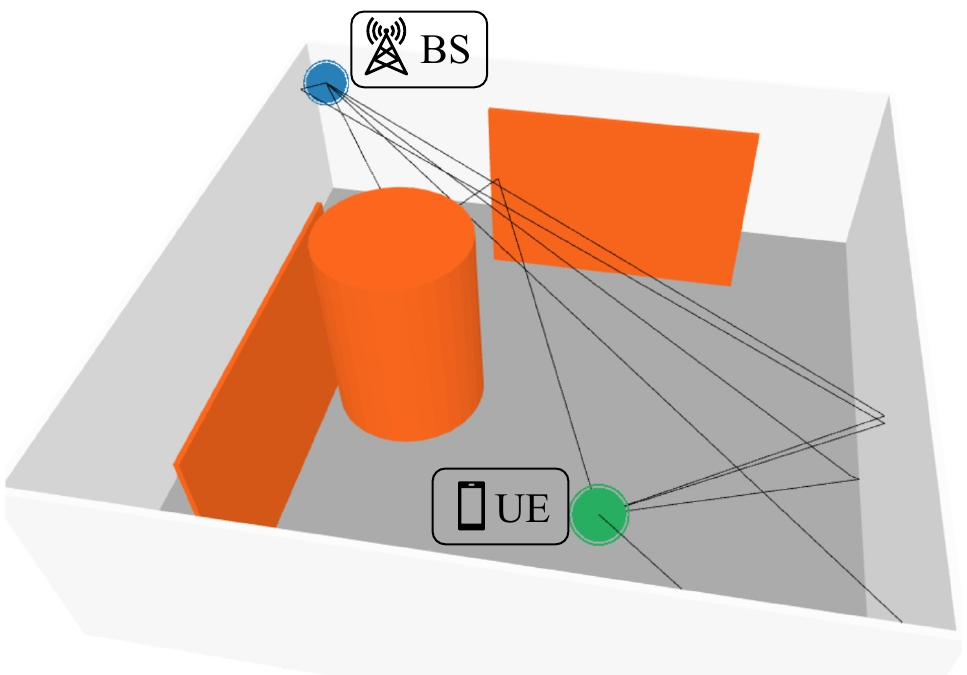}
        \caption{Illustration of generated scenes and paths simulated by Sionna.}
    \vspace{-3mm}
	\label{pic_sionna}
\end{figure}

To enable an environment-aware FM, environmental information should be included in the dataset.
In the constructed dataset, we provide three data formats for a specific scenario, namely the \textbf{original 3D environment, scene graph, and discretized scene graph}.
Specifically, indoor 3D environments are modeled using Blender\footnote{https://www.blender.org/}, a popular tool for creating 3D indoor scenes. All scenes share the same overall dimensions, i.e. $10m \times 10m \times 3m$, where the materials of the floor and the wall are assumed to be concrete and brick, respectively.  
In the room, 0$\sim$3 internal marble walls and 0$\sim$2 marble cylinders are placed, whose length and distribution are randomized.
In Fig.~\ref{pic_sionna}, we illustrate a typical 3D scene generated by Blender, where the indoor scenario includes two internal walls and a cylinder.
Given the computational complexity and high dimensionality of direct 3D environment processing, we extract top-view representations from 3D scenes that preserve both structural layouts and channel characteristics of the scene, denoted as the scene graph.
The scene graph representation is further discretized in matrix format $100 \times 100$, which is denoted as the discretized scene graph\footnote{Since DL models typically require data of uniform dimensions, we employ uniformly sized scenes and scene graphs in this study. However, it is important to note that our proposed method is applicable to scenes of varying sizes. In such cases, top-view maps are converted into square pixel images of the same size without
altering their spatial characteristics~\cite{10971878}. Specifically, maps of different physical extents are first padded to form square shapes, then rescaled into uniform-size pixel images.}.
We assign explicit physical meanings to the matrix elements: 1 indicates obstacle locations while 0 represents open space areas.

\subsubsection{Multi-task dataset}
Sionna~\cite{sionna}, an open-source library for link-level simulations based on TensorFlow, is employed to obtain datasets for multiple tasks in constructed environments. 
In particular, the BS is randomly placed at $(\pm4.75m,\pm4.75m,2.5m)$, and the element in the scene graph matrix is set as 2 to label the BS position.
Besides, the users are randomly distributed in the scenario. Then the Sionna ray tracing simulator~\cite{sionnart} is implemented to generate multipath channels in 3D environments, accurately representing propagation conditions by accounting for reflection, diffraction, and scattering.
The simulated paths between the BS and the UE are also illustrated in Fig.~\ref{pic_sionna}.
The parameters of the system are listed in Table~\ref{tab:parameters}.
Based on the generated channels, we can calculate the data samples for different tasks in the MIMO-OFDM transceiver, including the received signal, user positions, channels, etc.

\begin{table}[t]
\centering
\caption{System parameters during dataset generation.}
\label{tab:parameters}
\begin{tabular}{lc}
\toprule
\textbf{Parameter} & \textbf{Value} \\
\midrule
    BS antenna number $N_t$ & $64 = 16 \times 4$  \\
    UE antenna number & 1  \\
    Subcarriers $M$ & 48 \\
    Center frequency & 28 GHz \\
    Subcarrier spacing & 1.8 kHz \\
    UE height & 1 m \\
    User number $K$ & 4 \\
\bottomrule
\end{tabular}
\vspace{-3mm}
\end{table}

\vspace{-3mm}
\subsection{Simulation Setup and Training Details}
For hyperparameters in network training, we set the batch size as 100, and the initialized learning rate as 0.0001 with a cosine decay scheduler. We utilize Adam optimizer for model training with $ \rm betas=(0.9, 0.999)$.
The model undergoes training for 500 epochs over the multi-task dataset. 
The smallest version of GPT-2~\cite{gpt} with feature dimension $D = 768$ is adopted as the backbone, where all parameters are trained to obtain a FM dedicated for wireless communications.
All training and inference of the proposed model is conducted on four NVIDIA GeForce RTX 4090 24GB GPUs.
To validate the superiority of the proposed method, several baselines are implemented which can be categorized into the following groups.
To enhance the model's ability to process scene graphs of diverse sizes and resolutions, we apply data augmentation by resizing scene graphs to multiple scales during training.

\textbf{Single-task small models.} This class of methods
employ conventional DL architectures, such as MLPs, CNNs, and transformers, to address different tasks individually, which often have a relatively small number of parameters.
\begin{itemize}
    \item \textbf{MLP}: MLP is a typical architecture applied to various wireless communication problems~\cite{DECDNN,locdnn}. 
    \item \textbf{CNN}: Convolutional networks (CNNs) are suitable for tasks with high-dimensional wireless data~\cite{precnn}, such as the MIMO-OFDM channel matrix.
    \item \textbf{Transformer}: Transformer~\cite{Informer,ECCT}, which is also the basic block for FMs, exhibits excellent performance for wireless tasks, leveraging the attention mechanism. We implement the transformer-based model as~\cite{ECCT}.
\end{itemize}

\textbf{Dedicated methods for specific tasks.} The aforementioned typical DL architectures may not completely cover SOTA methods for all tasks. Thus, we have supplemented the baselines with additional effective approaches specific to each task, including both model-based methods and DL-based methods.
\begin{itemize}
    \item \textit{For multi-user precoding}, the eigen-based zero-forcing (ZF) algorithm~\cite{prezf} is a computationally efficient approach; while the WMMSE algorithm~\cite{WMMSE} is one of the most effective iterative algorithms. 
    \item \textit{For channel decoding}, the combination of the Belief Propagation (BP) model with hyper-graph network, namely hyper BP, is proposed in~\cite{BPdecoder} to achieve effective channel decoding. 
    We employ this method for 50 iterations, corresponding to a 100-layer neural network. 
    \item \textit{For MIMO detection}, ZF and linear minimum mean-squared error (LMMSE) detectors are classical methods with low complexity. In addition, DetNet in~\cite{detnet} and OAMP-Net~\cite{OAMP} are popular model-driven deep learning networks for this task.
    \item \textit{For channel estimation}, least squares (LS) constitutes a basic method.
\end{itemize}

\textbf{Single-task FMs.} These methods typically finetune FMs for a single task, which leverage the powerful modeling capabilities of FMs to improve the performance.  
\begin{itemize}
    \item \textbf{Task-specific FM}: Following the main idea in existing work~\cite{LLM4CP}, we train the FM for a single task. The task-specific FM directly employs the FM backbone in our proposed model with elaborately designed data encoders and decoders for different tasks.
    \item \textbf{Environment-aware task-specific FM}: To better evaluate the impact of involving environmental information, we also train an environment-aware task-specific FM that takes the embeddings of scene graphs as parts of inputs.
\end{itemize}

\textbf{Multi-task FMs.} 
We compare aforementioned initial attempts of multi-task FMs with the proposed method.
\begin{itemize}
    \item \textbf{Multi-task FM in~\cite{MTLLM}}: The authors in~\cite{MTLLM} exploit the multitasking ability of FM and propose a multi-task FM to unify multiple tasks with task-specific encoder-decoder pairs and a shared FM backbone.
\end{itemize}

Furthermore, we elaborate on the evaluation metrics for the tasks. The sum rate of users, which is the objective of multi-user precoding, is utilized as performance metric to evaluate the multi-user precoding task. NMSE serves as a crucial indicator for assessing channel estimation and MIMO detection precision. For channel decoding, BER is utilized; while for user localization, normalized error distance is employed.

\vspace{-3mm}
\subsection{Performance Evaluation}

\subsubsection{Environment awareness of the proposed model}

\begin{table*}
\centering
\caption{Performance of task-specific and multi-task FM with/without environment information.}
\label{tab:env}
\begin{tabular}{c|c|c|c|c|c}
\toprule
\multirow{2}{*}{\textbf{Method}} & \textbf{CE} & \textbf{Precoding} & \textbf{Det} & \textbf{Decoding} & \textbf{Loc}\\
& (NMSE $\downarrow$) & (Rate $\uparrow$) &  (NMSE $\downarrow$) &  ($\ln$ BER $\downarrow$) & (Error Dis $\downarrow$) \\
\midrule
FM(s) & 0.1387 & 19.44 & 0.001829 & -10.34 & 0.1516 \\
Env-aware FM(s) & \textbf{0.0982} & \textbf{19.71} & \textbf{0.001632} & \textbf{-10.54} & \textbf{0.1233} \\
\midrule
Improvements & 1.50 dB & 1.38\% & 0.49 dB & 0.20 dB & 18.67\% \\
\bottomrule
\end{tabular}
\vspace{-3mm}
\end{table*}

First, we evaluate the efficacy of incorporating environmental information in various tasks. To be specific, we compare the proposed environment-aware task-specific FM (namely ``Env-aware FM(s)”) with task-specific but environment-agnostic FM (namely ``FM(s)''). 
The simulation results are summarized in Table~\ref{tab:env}, where the SNRs for the received signal and the estimated channel are set as 10 dB; the pilot lengths for channel estimation and user localization are set as 4 and 8, respectively. Besides, we employ (64,32) polar codes in the channel decoding task. 

Given the strong correlation between the environment and the channel as discussed in Section~\ref{sec3-b}, the involvement of environmental information can provide prior knowledge of channel distributions. It is quite beneficial for channel estimation and user localization, the objectives of which are directly acquiring the CSI or positioning the users based on CSI. 
According to the results, the proposed environment-aware task-specific FM achieves better performance than task-specific FM for channel estimation and user localization. 
Specifically, the incorporation of environmental information reduces the NMSE of channel estimation by 1.50 dB and improves the user localization accuracy by 18.67\%.
For multi-user precoding and MIMO detection which take CSI as input, the environment-aware FMs also slightly outperform the environment-agnostic ones. It may be attributed to the scene graph to provide side information for noisy CSI, thus obtaining more accurate CSI for feature extraction.
Besides, since channel decoding is relatively independent of channel distributions, the two methods achieve comparable performance.
As shown in Table~\ref{tab:env}, the natural logarithm of BER achieved by environment-aware FM and environment-agnostic FM are -10.54 and -10.34, respectively.

Based on the analysis above, we summarize the main findings as follows.
\begin{itemize}
    \item Leveraging the multi-modal processing ability of FMs, the involvement of the scene graph as a part of input is vital to improve the performance and facilitate constructing an environment-aware wireless FM.
    \item The tasks that are highly related to the channel may benefit more from being informed of the prior knowledge extracted from environmental information.
\end{itemize}

\subsubsection{Task-specific performance analysis}
\begin{figure*}
	\centering 
	\subfigure[]{
		\includegraphics[width=0.32 \linewidth]{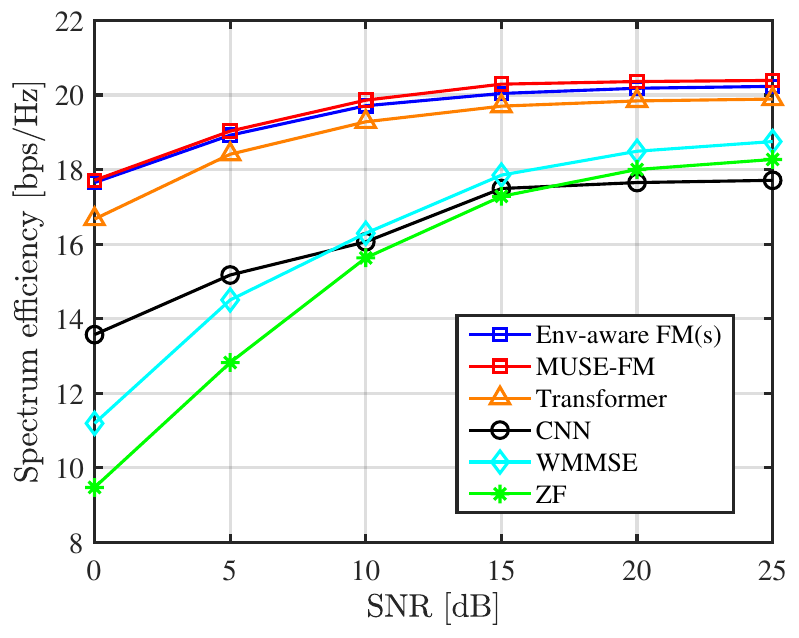}} 
        \subfigure[]{
		\includegraphics[width=0.32 \linewidth]{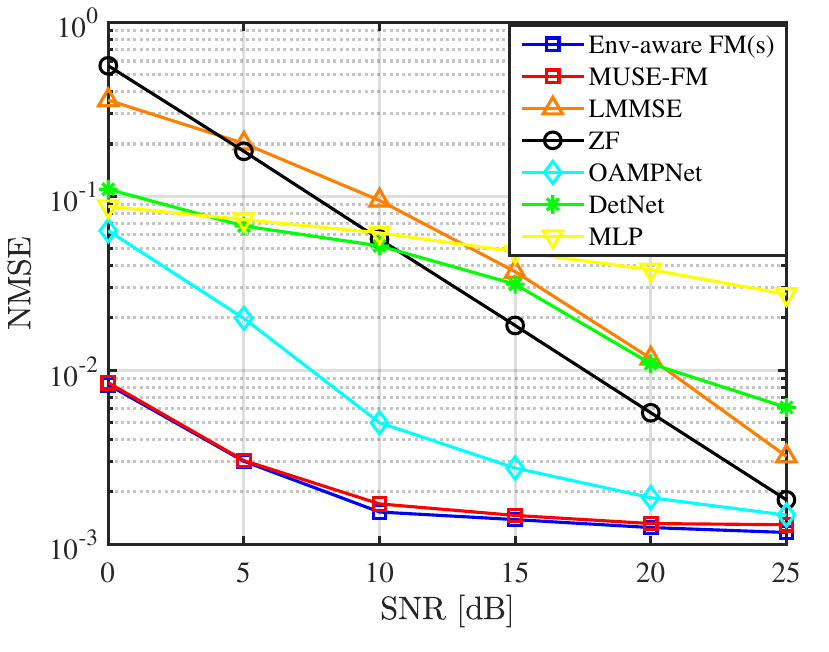}}
	\subfigure[]{
		\includegraphics[width=0.32 \linewidth]{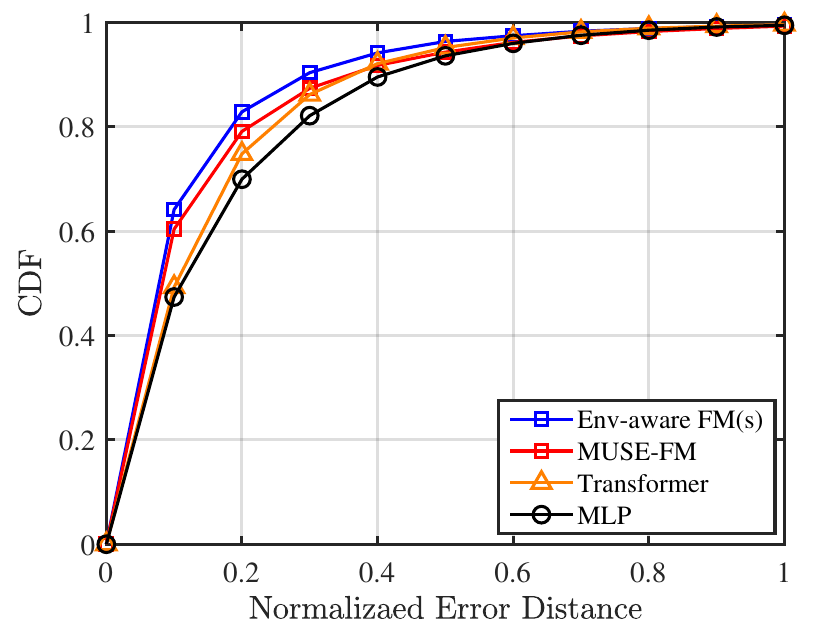}}
	\caption{Task-specific performance for: (a) Sum rate of multi-user precoding under different SNRs; (b) NMSE performance of MIMO detection under different SNRs; (c) CDF of normalized error distance for user localization.}
	\label{pic_taskperform}
    \vspace{-3mm}
\end{figure*}

For multi-user precoding, we compare the proposed method with both model-driven methods (ZF~\cite{prezf} and WMMSE~\cite{WMMSE}) and data-driven methods (CNN~\cite{precnn} and transformer-based method).
The sum rate performance is illustrated in Fig.~\ref{pic_taskperform} (a), where the multi-user channel is corrupted by noise with SNR values increasing from 0 dB to 25 dB.
It is observed that the ZF, WMMSE, and CNN-based methods fail to achieve satisfactory performance.
The proposed MUSE-FM significantly improves the sum rate, and outperforms shallow transformer and environment-aware task-specific FM-based methods with its strong feature extraction and joint learning capability. The advantage is more evident in the low SNR region, indicating the robustness of FM-based methods.

\begin{table}
\small
\centering
\caption{NMSE performance (in dB) versus SNR for channel estimation.}
\label{tab:CE}
\begin{tabular}{c|c|c|c|c|c}
\toprule
 {SNR(dB)} & \textbf{0} & \textbf{5} & \textbf{10}& \textbf{15} & \textbf{20}\\
\midrule
LS & -2.69 & -3.97 & -4.46 & -5.31 & -6.21 \\
ReEsNet~\cite{cecnn} & -5.68 & -6.08 & -6.39 & -6.62 & -6.84 \\
CE-ViT~\cite{cevit} & -6.78 & -8.46 & -9.00 & -9.65 & -10.11 \\
Env-aware FM(s) & \textbf{-9.35} & \textbf{-9.91} & \textbf{-10.08} & \textbf{-10.87} &  \textbf{-11.41} \\
MUSE-FM & -8.95 & -9.35 & -9.48 &  -9.87 &  -10.45 \\
\bottomrule
\end{tabular}
\vspace{-3mm}
\end{table}

In Fig.~\ref{pic_taskperform} (b), the NMSE performance of the proposed MUSE-FM on MIMO detection under different SNRs is presented, where the SNR increases from 0 dB to 25 dB. We also compare the performance with several baselines. 
As illustrated in Fig.~\ref{pic_taskperform} (b), despite the low computational complexity, the ZF and LMMSE-based methods show relatively high NMSE performance in low SNR regime.
To solve the problem, DL-driven methods, including OAMPNet~\cite{OAMP}, DetNet~\cite{detnet} and MLP-based~\cite{DECDNN} method, are proposed to efficiently improve the recover accuracy. 
In addition, the proposed FMs with environmental awareness further improve NMSE performance and outperform baselines, especially in the low SNR regime.
Moreover, we observe that the exploitation of statistical information of noise (i.e. SNR) as prior knowledge helps improve the performance. It is validated by the superior performance of OAMPNet and proposed MUSE-FM over others, which utilize the noise variance in the model.

Next, we discuss the performance of user localization.
The cumulative distribution function (CDF) curve of the normalized positioning error distance is presented in Fig.~\ref{pic_taskperform} (c).
The SNR of the received data is set as 10 dB. The proposed environment-aware task-specific FM achieves the most accurate user localization, while the proposed MUSE-FM obtains comparable performance with marginal gap. They outperform the traditional transformer-based method and the simple low-cost MLP-based method~\cite{locdnn}. Specifically, the ratio of samples with normalized error distance less than 0.1 is 64.13\% for the proposed environment-aware task-specific FM and 60.38\% for the proposed MUSE-FM. The ratio drops to 49.27\% for the traditional transformer-based method and 45.37\% for the MLP-based method, validating the advantage of the proposed method.

\begin{table}
\centering
\caption{Comparison of the negative natural logarithm of BER for different $E_b/N_0$ in channel decoding task.}
\label{tab:cc}
\begin{tabular}{c|c|c|c}
\toprule
 { $\bf{E_b/N_0}$} & \textbf{4} & \textbf{5} & \textbf{6} \\
\midrule
Hyper BP~\cite{BPdecoder} & 4.59 & 6.10 &  7.69 \\
ECCT~\cite{ECCT} & 6.99 & 9.44 & 12.32 \\
Env-aware FM(s) & \textbf{7.68} & \textbf{10.54} & \textbf{13.82} \\
MUSE-FM & 7.56 & 10.38 & 13.72 \\
\bottomrule
\end{tabular}
\vspace{-4mm}
\end{table}

Table.~\ref{tab:CE} shows the NMSE performance versus SNR for channel estimation, where the SNR increases from 0 dB to 20 dB. Besides the traditional LS method, we also compare our method with the CNN-based method, ReEsNet~\cite{cecnn}, and transformer-based method, CE-ViT~\cite{cevit}. 
LS performs the worst as it ignores the effect of the noise. 
The performance of ReEsNet and CE-ViT exceeds that of LS, but still requires further improvements.
The proposed FMs, both task-specific and multi-task, outperform other methods and maintain accurate estimation in low SNR regime.

\begin{table*}
\centering
\caption{Evaluation of performance of different tasks trained with different sizes of dataset.}
\label{tab:few}
\begin{tabular}{c|c|c|c|c|c|c|c|c|c|c}
\toprule
\multirow{2}{*}{\textbf{Data size}} & \multicolumn{2}{c}{\textbf{CE}} &  \multicolumn{2}{c}{\textbf{Precoding}} & \multicolumn{2}{c}{\textbf{Det}} &  \multicolumn{2}{c}{\textbf{Decoding}}  & \multicolumn{2}{c}{\textbf{loc}}\\
& \multicolumn{2}{c}{(NMSE $\downarrow$)} & \multicolumn{2}{c}{(Rate $\uparrow$)} &  \multicolumn{2}{c}{(NMSE $\downarrow$)}  &  \multicolumn{2}{c}{($\ln$ BER $\downarrow$)}   &  \multicolumn{2}{c}{(Error Dis $\downarrow$)}  \\
\midrule
Ratio & FM(s) &  FM(m) &  FM(s) & FM(m) &  FM(s) & FM(m) & FM(s) & FM(m) &  FM(s) &  FM(m)\\
\midrule
2\% dataset & 0.2353 & \textbf{0.2056} & 13.73 & \textbf{14.24} & 0.029454 & \textbf{0.023825} & -7.69 & \textbf{-7.91} & 0.3150 & \textbf{0.2817} \\
10\% dataset & 0.1526 & \textbf{0.1448} & 17.28 & \textbf{17.80} & \textbf{0.007424} & 0.007726 & -8.08 & \textbf{-8.18} & 0.2052 & \textbf{0.1938}\\
20\% dataset & 0.1299 & \textbf{0.1272} & 18.36 & \textbf{18.68} & \textbf{0.003105} & 0.003218 & -8.23 & \textbf{-8.63} & 0.1740 & \textbf{0.1723} \\
50\% dataset & \textbf{0.1126} & 0.1180 & 19.20 & \textbf{19.49} & \textbf{0.001888} & 0.002097 & -9.37 & \textbf{-9.56} & \textbf{0.1422} & 0.1505 \\
80\% dataset & \textbf{0.1044} & 0.1164 & 19.59 & \textbf{19.75} & \textbf{0.001611} & 0.001702 & -9.92 & \textbf{-10.09} & \textbf{0.1284} &  0.1481 \\
full dataset  & \textbf{0.0982} & 0.1125 & 19.71 & \textbf{19.76} & \textbf{0.001632} & 0.001736 & \textbf{-10.54} & -10.38  & \textbf{0.1233} & 0.1415\\
\bottomrule
\end{tabular}
\vspace{-3mm}
\end{table*}

Finally, we focus on the performance of channel decoding.
The results are reported in Tab.~\ref{tab:cc} where we present the negative natural logarithm of the BER under different normalized SNR (i.e. $E_b/N_0$). 
For transformer-based channel decoding, namely ECCT, we select the largest version presented in~\cite{ECCT}.
We observe that our approach significantly outperforms the current methods for all different SNR values.
Despite the similar structure to ECCT, equipped with larger model size, the proposed model further improves the performance. 

From the simulation of all selected tasks, we conclude that:
\vspace{-3mm}
\begin{itemize}
    \item The proposed environment-aware FMs, trained either on a single task or multiple tasks, outperform existing baselines for all tasks, indicating superior feature extraction capabilities of FMs and their potential in wireless domain.
    \item The advantages of FM-based models are more evident in challenging scenarios, such as the low SNR regime, which may be attributed to the robustness and generalization ability of FMs.
    \item With the multi-task learning ability of FMs, the proposed MUSE-FM achieves comparable performance with environment-aware task-specific FM.
\end{itemize}
Moreover, MUSE-FM demonstrates advantages in training and inference overhead compared to task-specific FMs.
Regarding memory overhead, with 134.44M parameters, MUSE-FM incurs significantly lower storage costs compared to deploying five task-specific FMs (each 93.20M parameters, totaling 466M parameters).
Compared to the sequential training and inference of multiple task-specific FMs, MUSE-FM reduces training latency by 44.3\% and inference latency by 48.9\%.

\subsubsection{Scaling behavior}
We next investigate how the model performs with different sizes of dataset. It is crucial for efficient deployment of DL-based models, which determines the trade-off between performance and the cost of data collection and network training.
In this part, we evaluate the scaling behavior of the proposed MUSE-FM, compared with environment-aware task-specific FMs, as shown in Table~\ref{tab:few}. Note that both task-specific FMs and multi-task FMs incorporate environmental information, while the notations ``FM(s)'' and ``FM(m)'' are shorthand for ``Env-aware FM(s)'' and ``MUSE-FM'', respectively, used for table conciseness.
Besides training the model on the full dataset, we train the model with 2\%, 10\%, 20\%, 50\% and 80\% of the dataset, respectively (i.e., we utilize 1, 5, 10, 25, and 40 samples for each training scenario).
It is observed that for each task, as the data volume increases, the performance of both task-specific FM and multi-task FM gradually improves and asymptotically approaches their performance ceiling.

\begin{table*}
\centering
\caption{Performance of FM with proposed FM with unified encoder-decoder pair and FM with task-specific encoder-decoder pairs.}
\label{tab:endecoder}
\begin{tabular}{c|c|c|c|c|c}
\toprule
\multirow{2}{*}{\textbf{Method}} & \textbf{CE} & \textbf{Precoding} & \textbf{Det} & \textbf{Decoding} & \textbf{Loc}\\
& (NMSE $\downarrow$) & (Rate $\uparrow$) &  (NMSE $\downarrow$) &  ($\ln$ BER $\downarrow$) & (Error Dis $\downarrow$) \\
\midrule
FM with task-specific encoder-decoder pairs~\cite{MTLLM} & 0.1183 & 19.50 & 0.001778 & -10.30 & \textbf{0.1389} \\
MUSE-FM & \textbf{0.1125} & \textbf{19.76} & \textbf{0.001736}  & \textbf{-10.38} & 0.1415\\
\bottomrule
\end{tabular}
\vspace{-4mm}
\end{table*}

When trained on the full dataset, the environment-aware task-specific FM slightly outperforms MUSE-FM, benefiting from avoiding inter-task compromises and the higher complexity of multi-task training.
As the dataset size decreases, the MUSE-FMs trained on 20\% and 50\% data samples achieve comparable performance with corresponding task-specific FMs; while MUSE-FMs even outperform task-specific ones in some tasks (multi-user precoding for instance).
Furthermore, as shown in Table~\ref{tab:few}, when data samples are extremely limited (only 2\% and 10\% of the dataset is accessible), the proposed MUSE-FM outperforms the task-specific FM on most tasks. For example, MUSE-FM achieves 17.80 bps/Hz for multi-user precoding, versus 17.24 bps/Hz for the task-specific FM, when trained on 10\% of the dataset. When trained with only 2\% dataset, MUSE-FM improves the performance for all selected tasks, compared with task-specific FMs.
The main reason may be analyzed as follows:
with limited data samples, the task-specific FM, accessing only one task's dataset, is more prone to overfitting, leading to performance degradation. In contrast, the proposed MUSE-FM leverages its multi-task learning capability to jointly learn from multiple datasets, enhancing joint feature representation. 

It should be noted that, due to the large size of the constructed dataset, our proposed method trains all FM parameters to build a wireless communication-specific FM based on pre-trained parameters. 
For fair and direct comparison, we also train all FM parameters when training with only part of the dataset.
Consequently, it results in overfitting and performance degradation even for multi-task FM. 
In practice, when data samples are scarce, we can train only a small portion of the parameters in the FMs to avoid overfitting and improve the performance.

We summarize the main insights for the experiments in this subsection as follows.
\vspace{-1mm}
\begin{itemize}
    \item Since FMs possess the capability to learn complex patterns from vast data, gradually increasing data volume enhances the performance, both for multi-task FM and task-specific FM.
    \item When data volume is sufficient, task-specific FMs achieve superior performance; however, with limited data, multi-task models benefit from joint learning across multiple tasks and datasets. This illustrates the capability of multi-task FMs to exploit different types of datasets of wireless communications.
\end{itemize}
\vspace{-1mm}

\begin{figure}
	\centering 
	\includegraphics[width= 0.83 \linewidth]{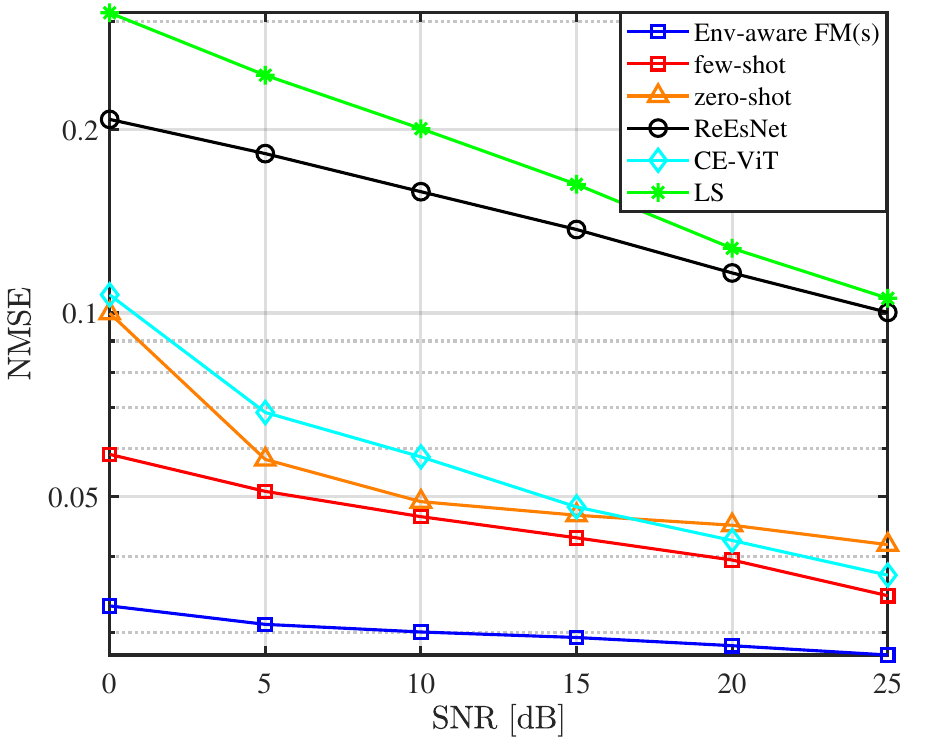}
        \caption{NMSE performance for channel estimation with pilot length =6.}
\vspace{-3mm}
	\label{pic_ce6}
\vspace{-3mm}
\end{figure}

\subsubsection{Impact of prompt-guided unified data encoder-decoder pair}
In this subsection, the effectiveness of the proposed prompt-guided unified data encoder-decoder pair is evaluated.
As discussed in Section~\ref{sec3-c}, designing separate data encoders and decoders as~\cite{MTLLM} for each task and parameter helps task alignment and facilitates multi-task learning, but it incurs high computational and deployment overhead, given the diverse tasks and parameter variations. 
Besides, when encountering a new task or parameter configuration with different input/output formats, FMs with task-specific data encoder-decoder pairs may not work. Instead, a new task-specific encoder-decoder pair is required to be designed and trained, which limits the scalability of the models.
Therefore, we propose a unified prompt-guided data encoder-decoder pair to improve the scalability of multi-task FM, while retaining superior performance for all tasks.
To validate the effectiveness, we compare the proposed MUSE-FM with multi-task FM with task-specific encoder-decoder pairs in~\cite{MTLLM}.

The simulation results for the proposed MUSE-FM and that with task-specific encoder-decoder pairs are provided in Table~\ref{tab:endecoder}. The SNRs for the received signal and estimated channel are set as 10 dB; the pilot lengths for channel estimation and user localization are set as 4 and 8, respectively.
Although only a unified encoder-decoder pair is employed, 
the proposed method achieves comparable performance with multi-task FM with task-specific encoder-decoder pairs in~\cite{MTLLM}.
It even slightly outperforms multi-task FM in~\cite{MTLLM} for channel estimation, multi-user precoding and MIMO detection.
Its comparable performance mainly stems from two reasons, involving domain knowledge and learning joint feature representation. Specifically, the parameters of the prompt-guided unified data encoder-decoder pair are dynamically generated by task-specific instructions, which contain critical domain knowledge and key parameters. With extra information included, performance is improved.
Besides, the proposed unified encoder-decoder pair extracts joint representation from various tasks, leveraging the relations among the tasks to achieve better performance.

In conclusion, we observe that the proposed prompt-guided unified data encoder-decoder pair achieves comparable performance with task-specific encoder-decoder pairs, while improves the adaptability of the multi-task FM.

\subsubsection{Generalization ability}

Then we evaluate the generalization ability of MUSE-FM for a new parameter configuration (e.g. antenna number, pilot length, user number).
Fig.~\ref{pic_ce6} utilizes channel estimation for instance to illustrate the ability of the proposed MUSE-FM in handling new task configurations.
During the training process, the pilot length is set as 4, while the pilot length is changed to 6 during evaluation.
Varying pilot quantities alter input dimensions, rendering FM with task-specific encoder-decoder pairs inapplicable. In contrast, our proposed method remains directly applicable to new parameter configurations. To comprehensively demonstrate its capability, except the existing baselines, we compare: (1) FM trained specifically for pilot length = 6, (2) Zero-shot performance where FM trained with pilot length = 4 is applied to pilot length = 6, and (3) Few-shot performance where FM trained with pilot length = 4 is fine-tuned with 1\% data for 5 epochs.
As shown in Fig.~\ref{pic_ce6}, the zero-shot performance of our proposed method surpasses ReEsNet and LS-based methods and achieves comparable performance with CE-ViT. With only minimal fine-tuning (1\% data for 5 epochs), our method outperforms all baselines, indicating the generalization ability of the proposed MUSE-FM .

\section{Conclusions} \label{sec-con}
\vspace{-1mm} 
In this work, we propose a multi-task environment-aware FM, namely MUSE-FM for wireless communications with a unified architecture to handle multiple tasks and incorporate scenario information.
Through multi-modal input integration of environmental context and wireless data, the proposed model manages to facilitate cross-scenario feature extraction and enhances the environmental adaptability. 
The proposed prompt-guided data encoder-decoder pair employs instructions to dynamically generate task-specific parameters, and thus handles data with heterogeneous formats and distributions across multiple tasks with a unified encoder-decoder pair.
Simulation results have demonstrated the multi-task learning and generalization capabilities of the proposed method.
Besides, incorporating environmental information facilitates cross-scenario learning ability, while the proposed prompt-guided unified encoder-decoder pair improves the scalability of the model.
Furthermore, with limited data samples, MUSE-FM harnesses cross-task dependencies for knowledge transfer and enhance joint optimization compared with task-specific FMs.
Future works can focus on more effective use of environmental information.
In addition, the end-to-end optimization for the transceiver can be used to further improve the performance.

\footnotesize

\vspace{-2mm} 
\bibliographystyle{IEEEtran}
\bibliography{reference, IEEEabrv}

\normalsize

\end{document}